# Chapter 14

# Injection and Dumping Systems

*B. Goddard, A. Lechner and J. Uythoven*\*

CERN, Accelerator & Technology Sector, Geneva, Switzerland

## 14 Injection and dumping systems

### 14.1 Injection and dumping systems overview

The beam transfer into the LHC is achieved by the two long transfer lines TI2 and TI8, together with the septum and injection kicker systems, plus associated machine protection systems to ensure protection of the LHC elements in case of a mis-steered beam. The LHC is filled by approximately 10 injections per beam. The MKI kicker pulse length consists of an 8 μs flat-top, with a rise time of 0.9 μs and a fall time of 2.5 μs. Filling each ring takes 8 minutes with the SPS supplying interleaved beams to other facilities. The foreseen increase in injected intensity and brightness for the HL-LHC means that the protection functionality of the beam-intercepting devices needs upgrading, see Ref. [1]. In addition, the higher beam current significantly increases the beam-induced power deposited in many elements, including the injection kicker magnets in the LHC ring.

The beam dumping system is also based on DC septa and fast kickers, with various beam intercepting protection devices including the beam dump block. Again, the significant change in the beam parameters for the HL-LHC implies redesign of several of the dump system devices, because of the increased energy deposition in the case of direct impact, but also because of increased radiation background that could affect the reliability of this key machine protection system [1].

In the following sections the changes planned in the light of the HL-LHC for the different LHC beam transfer systems are described.

### 14.2 Injection systems

The high injected beam intensity and energy mean that precautions must be taken against damage and quenches, by means of collimators placed close to the beam in the injection regions. The layout of the injection region and associated protection devices is shown schematically in Figure 14-1. The beam to be injected passes through five horizontally deflecting steel septum magnets (MSI) with a total deflection of 12 mrad, and four vertically deflecting kickers (MKI) with a nominal total kick strength of 0.85 mrad. Uncontrolled beam loss resulting from errors (missing, partial, badly synchronized, or wrong kick strength) in the MKI could result in serious damage to downstream equipment in the LHC injection regions, in particular the superconducting separation dipole D1, the triplet quadrupole magnets near ALICE or LHCb experiments, or in the arcs of the LHC machine itself. Damaging detector components, in particular those close to the beam pipe, with excessive showers generated by lost protons, is also possible.

---

\* Corresponding author: Jan.Uythoven@cern.ch

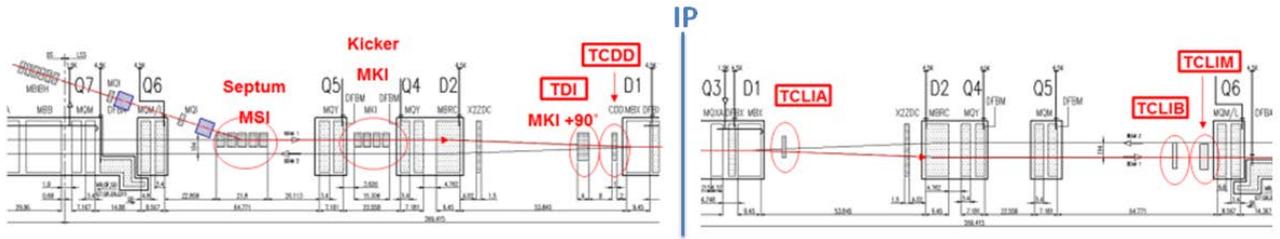

Figure 14-1: Overview of the present injection system into the LHC and the associated protection devices (Beam 1, IR2). The beam is injected from the left hand side.

### 14.2.1 Injection absorber (TDIS)

The TDI is a movable two-sided absorber. Its purpose is to protect machine elements in case of MKI malfunctions and to intercept low-intensity bunches during set-up or commissioning of the injection system. The present TDI, see Ref. [2], needs to be adapted to HL-LHC and LIU beam parameters. This implies a change of absorber materials but no significant change to the total absorber length. In addition, problems with the present design have become apparent during the first years of LHC operation. Instead of having one long jaw (4.185 m) as in the present TDI, the new TDIS (where the S stands for Segmented) is foreseen to comprise shorter absorbers accommodated in separate tanks, which would imply an increase in the total length.

The protection objective stated above concerns both mis-deflections of the injected beam due to MKI faults as well as accidental kicks of the stored beam due to MKI timing errors. In the first of these cases, too little or zero kick strength causes the incoming beam to impact the upper TDIS jaws. In the second case, the miskicked stored beam impacts the lower TDIS jaws. The protection objectives should be met for any impact condition, including cases where beams are swept over the aperture (misfires during MKI rise time) or where a full batch is grazing the TDIS jaws due to a non-nominal MKI kick strength. The maximum possible impact parameter (vertical distance between equilibrium orbit and impact position) on the absorber front face is estimated to be 3 cm. The largest energy deposition in downstream magnets and the highest stresses in the jaw itself are expected for small impact parameters (around 1–2 $\sigma$, which is about 1 mm from the absorber edge), which could occur in the case of a magnet breakdown. New low-$Z$ absorber materials, replacing boron nitride (BN5000, 1.93 g/cm$^3$) used for the present TDI, are being studied.

If the protection objectives stated above cannot be met for a full injected beam of 288 bunches, the total number of injected bunches will need to be decreased accordingly. This has to be done in conjunction with a reduction of the MKI pulse length in order to limit the number of circulating bunches that could be miskicked onto the TDIS in the case of a timing error.

Apart from the aperture requirements for circulating and injected beams that are identical to the present TDI, see Ref. [2], there are additional aperture requirements coming from the ALICE ZDC, see Ref [3].

The jaw positioning requirements are listed in Table 14-1. Jaws at fully closed positions will be required during optimization of the injection systems. A fully open position is important to reduce beam impedance, and thus reduce heating when not injecting beam. The positions given are relative to the closed orbit, which nominally includes the separation and crossing bumps.

Table 14-1: TDIS position relative to the orbit

| Characteristic | Unit | Value |
| --- | --- | --- |
| Position fully closed | mm | 1.0 |
| Typical position during injection | mm | ± 3.8 |
| Position when no injection | mm | ± 55.0 |

The beam impedance should be minimized during the injection process (absorber blocks close to the beam) and when injection is finished and the jaws are retracted (absorber blocks in parking position). The resulting beam impedance should be acceptable from the beam stability perspective and in terms of the deposited power in the TDIS. If necessary water cooling of the absorbers will need to be installed.

The correct positioning of the TDIS is a vital element of the machine protection system during injection. Redundant position measurement of the blocks is required with redundant interlock channels going to the beam interlock system (BIS). Both LVDT and laser interferometric position measurements are foreseen. A beam energy tracking system (BETS) to guarantee the correct position at injection energy is foreseen and will use the laser interferometric position readings. For a multi-module TDIS system the correct alignment with beam between the different modules will be very important, as is illustrated by Figure 14-2.

The new TDISs will be approximately at the same position as the present TDI, as they need to be at 90° betatron phase advance relative to the MKIs. The extension of the TDIs is to take place in the upstream direction, keeping the distance between the downstream end of the TDI and D1 unchanged. For IP2 this means that the large 800 mm chamber upstream of the TDI needs to be shortened. This has been discussed with the ALICE experiment and is acceptable as long as a minimum opening position is guaranteed when the jaws are retracted.

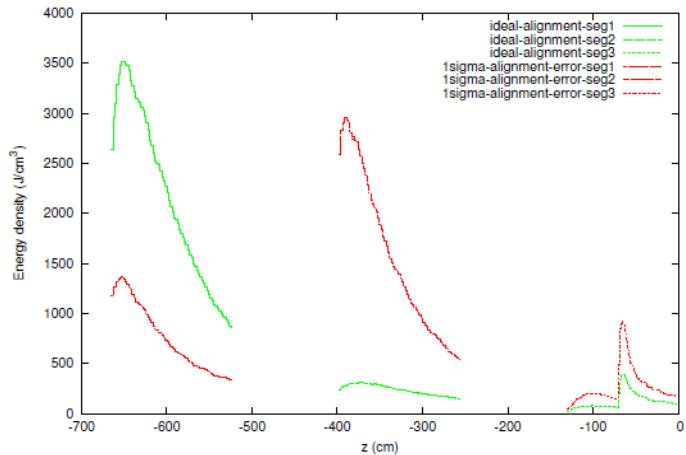

Figure 14-2: Energy deposition in a TDIS consisting of three modules (graphite, graphite and aluminium, and copper) for grazing beam impact and showing the effect of a 1 $\sigma$ alignment difference between the first and second module (courtesy of N. Shetty).

### 14.2.2 Auxiliary injection protection collimator (TCLIA)

The TDIS will be complemented by two auxiliary collimators, TCLIA and TCLIB, which are located on the other side of the IP, at phase advances that are designed to be ±20° (modulo 180°) relative to the TDI. The position of these auxiliary absorbers is shown in Figure 14-1. The TCLIA is a two-sided auxiliary injection protection collimator located between the separation dipoles in IR2 (righthand side) and IR8 (lefthand side). Owing to its position close to the D1, the TCLIA also hosts the other circulating beam. Together with another auxiliary collimator (TCLIB) it complements the primary injection protection absorber TDIS in case of injection kicker (MKI) failures. The present TCLIA consists of low-$Z$ absorber blocks (graphite R4550, 1.83 g/cm$^3$) with a total active length of 1 m. It is being evaluated whether the present TCLIA design, in particular the absorber material, can be retained for the HL-LHC era or if a new design is required due to the increased beam brightness. In addition, it is to be determined if the present TCLIA design is compatible with aperture requirements imposed by the ALICE ZDC [3] operation during heavy-ion physics runs. If the present aperture is found to be incompatible, a new design with a larger stroke for the open position will be required.

In the event of beam impact on the TCLIA, no damage must occur to downstream machine elements, in particular to the superconducting separation dipole D2. However, a magnet quench cannot be excluded from all MKI failure scenarios. In all cases, the TCLIA itself must not sustain damage during injection failures.

The TDI and TCLI elements need to protect the LHC arc aperture from mis-steered beams. Tracking studies showed that settings of 6.8 $\sigma$ for the TDI and TCLI systems adequately protect the LHC arc aperture against MKI flashovers, see Ref. [4]. This result depends on the injected intensity, and it may be necessary to reduce the 6.8 $\sigma$ settings slightly for injection of higher intensity beams – this needs to be analyzed in the context of the beam cleaning collimation system settings for injection.

### 14.2.3 Auxiliary injection protection collimator [TCLIB]

The TCLIB is a two-sided auxiliary injection protection collimator located between the matching section quadrupoles Q5 and Q6 in IR2 (righthand side) and IR8 (lefthand side). Together with the auxiliary collimator TCLIA that has already been described, it complements the primary injection protection absorber TDIS in case of injection kicker failures. The present TCLIB accommodates low-Z absorber blocks (CfC AC150, 1.67 g/cm$^3$) with a total active length of 1 m. It is complemented by a mask in front of Q6 (TCLIM), which is required to absorb secondary showers from the TCLIB such that damage to the Q6 and other downstream equipment is prevented. It is to be evaluated whether the present TCLIB design, in particular the absorber material, can be retained for the HL-LHC era or if a new design is required due to the increased beam brightness.

Depending on the injection kicker strength, badly injected beam will be absorbed by TDIS, TCLIA, or TCLIB. It is shown in Figure 14-3 that TCLIB intercepts up to 28% of the 5 $\sigma$ area of single particle emittance for injection kicker failures with ±10% kick nominal strength, which corresponds to a grazing incidence of the beam on the TDIS.

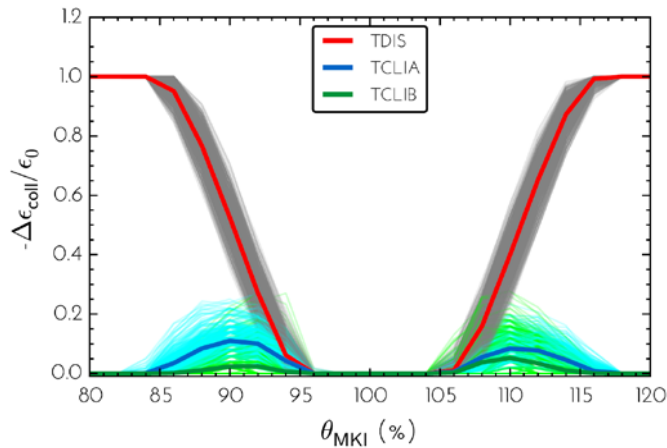

Figure 14-3: Fraction of 5 $\sigma$ single particle emittance absorbed by the different injection protection absorbers TDIS, TCLIA, or TCLIB for different injection kicker amplitudes (courtesy of F.M. Velotti).

### 14.2.4 Injection protection masks TCDD and TCDDM

TCDD and TCDDM are masks upstream of the superconducting D1 in IR2 (left) and IR8 (right), which complement the primary injection protection absorber TDIS in case of injection kicker malfunctions, see Figure 14-1. The masks are 1 m long Cu blocks, which in IR2 are required to open for ALICE ZDC data-taking during ion runs. The TCDD/TCDDM is required to sufficiently absorb secondary showers or scattered protons from the TDIS such that it prevents damage to D1 and other equipment for any possible TDIS impact conditions and for all HL-LHC and LIU beam parameters. In IR2, any new design or new mask position must account for ALICE ZDC aperture requirements. The aperture of the present TCDD/TCDDM might need to be further minimized to protect D1 against particle showers from the high intensity beams grazing the TDIS absorber. The aperture requirements for the circulating beam and damage limits of the D1 separation dipole are important parameters to be taken into account.

### 14.2.5 Injection protection mask (TCLIM)

The injection protection mask (TCLIM) is a fixed mask upstream of the superconducting matching section quadrupole Q6 in IR2 (righthand side) and IR8 (lefthand side). It is presently of 1 m active length and made of stainless steel. In the event of of injection kicker malfunctions leading to beam impact on the auxiliary injection protection collimator TCLIB, the mask is required to absorb secondary showers from the TCLIB such that it prevents damage to Q6 and other downstream equipment. Protection has to be ensured for all HL-LHC and LIU beam parameters. It is to be evaluated whether the present TCLIM design can be retained or if a new design is required, possibly with a different length and material.

### 14.2.6 Injection kicker magnet (MKI)

The injection kicker magnets MKI installed in IR2 and IR8, see Figure 14-1, deflect the injected beam onto the LHC closed orbits. During Run 1 of the LHC a number of issues have been encountered with the MKI magnets installed prior to LS1. These include beam-induced heating of the ferrite yoke, inefficient cooling of the ferrite yoke, electrical flashovers, beam losses due to macro particles falling into the beam, and electron cloud [5, 6].

If the ferrites of the injection kicker magnets reach a temperature above their Curie temperature their magnetic properties are compromised and the beam cannot be injected. Reducing beam-induced heating, additional cooling, and/or ferrites with a higher Curie temperature avoid waiting periods without beam before the beam can be injected into the LHC. A reduced magnetic field from the injection kickers is also a machine protection issue, possibly leading to quenches of downstream magnets. The beam impedance of the MKI magnets has been reduced in LS1 by completing the number of beam screen conductors to 24, see Figure 14-4. However it cannot be guaranteed that this will sufficiently reduce beam-induced heating for HL-LHC operation. Operational experience during Run 1 has shown that a beam power deposition of about 160 W/m occasionally limited the ability to inject beam. The present power deposition estimate for HL-LHC beams is 190 W/m, assuming use of the improved beam screen as installed in LS1. For this reason a prototype MKI magnet with additional cooling and different ferrite types will be developed for testing in the LHC. A ferrite such as CMD10, which has a higher Curie temperature than the CMD5005 or 8C11 presently used for the MKI yoke, would permit high intensity beam operation with better availability. However, operating at higher yoke temperatures will result in higher pressure in the vacuum tank, which may result in an increased electrical breakdown and surface flashover rate. Further optimization of the capacitively coupled end of the beam screen is being made to further reduce the electric field strength and so the likelihood of surface flashovers.

The exchange of an MKI during the third technical stop (TS3) of 2012, and its subsequent operation in the LHC, demonstrated that electron cloud in the ceramic tube can limit the beam intensity until conditioning had occurred (~250 hours of beam). Electron cloud in the ceramic tube results in a pressure increase and can therefore prevent operation of the injection kickers. A low SEY coating would eliminate multipactoring, and thus the related pressure rise, permitting more reliable operation of the injection kickers. Thus, research and development of special coatings for the inner surface of the ceramic tube is being carried out, and it is planned to include a coating on the tube of the prototype MKI.

After a comprehensive study programme in 2011, the macro particles causing beam losses around the MKIs were identified as fragments originating from the ceramic tube inside the MKI magnets [7]. Thus, the ceramic tube of MKI8D installed during TS3 was subjected to improved cleaning, which included iterations of flushing of the inside of the tube with $N_2$ at 10 bar and dust sampling, until no significant further reduction of macro particles was noted. Before TS3, MKI8D exhibited the highest rate of beam–dust particle interactions of all MKIs in P8; the replacement MKI8D, in operation after TS3, exhibited the lowest rate. Extensive additional cleaning was carried out on the ceramic tubes installed during LS1.

Modification of the series of MKI magnets is not part of the HL-LHC baseline. The necessity of the described changes for the series of magnets depends on the performance of the magnets installed in LS1 and the final beam parameters to be used for the HL-LHC beams (especially the bunch length). The installation of a prototype magnet to test the developed technologies is foreseen.

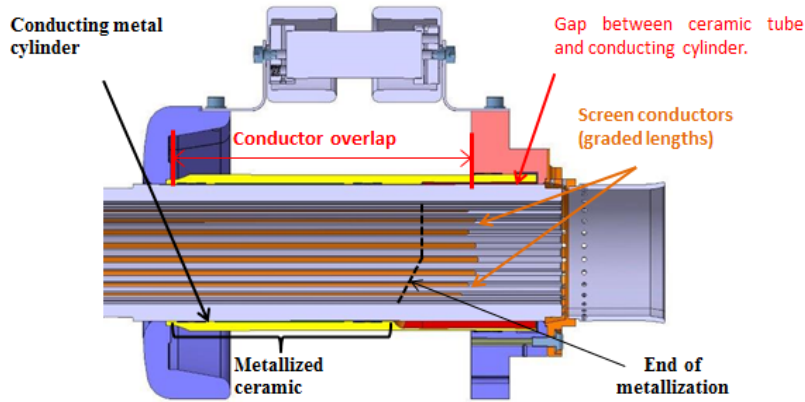

Figure 14-4: View of the capacitively coupled extremity of the MKI beam screen supporting the 24 screen conductors and the conductor overlap (courtesy of M.J. Barnes).

## 14.3 Beam dumping system

The beam in the LHC is aborted or dumped by a dedicated system based on pulsed extraction kickers and DC septum magnets located in the dedicated insertion in P6, followed by a dilution kicker system MKB, a long drift chamber, and a graphite beam dump absorber block (TDE) kept under $N_2$ gas at atmospheric pressure. The 3 μs rise time of the extraction kicker field is synchronized by a highly reliable timing system to a beam-free abort gap in the circulating bunch pattern. The horizontal and vertical dilution kickers are powered with anti-phase sinusoidal currents in order to paint the bunches onto the TDE with an elliptical shape, see Figure 14-5.

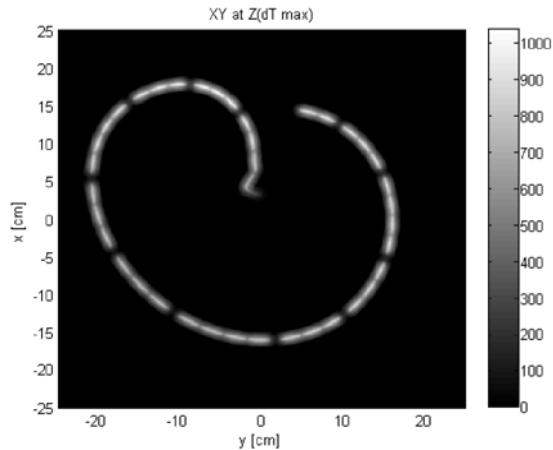

Figure 14-5: Sweep form of 25 ns spacing LHC beam on TDE dump block with a bunch intensity of $1.7 \times 10^{11}$ protons per bunch and the resulting temperature rise.

### 14.3.1 Beam dumping system limitations

The LHC beam dump block TDE and its entrance window [8, 9] will need to withstand the high intensity HL-LHC beams. Simulations are foreseen to verify if the present system can withstand the repeated beam impact of these beams. Concerns are the temperature rise of the TDE block, which could lead to a venting of nitrogen overpressure within the TDE. Another concern is the entrance window of the TDE. As a possible solution the installation of additional dilution kicker magnets on the beam dump lines TD62 and TD68 could be required.

As the performance mentioned above can possibly be met by the existing system, any modifications are not part of the present baseline.

### 14.3.2 Beam dumping system absorber (TCDS)

Several failure modes exist in the synchronization system and in the kicker switches that could lead to an asynchronous dump, in which the full beam intensity would be swept across the LHC aperture by the rising kicker field. Without dedicated protection devices this would lead to massive damage of the LHC magnets in LSS6 and the downstream arcs 5–6 and 6–7 and, depending on the operational configuration, a number of collimators and possibly experimental triplet magnets. The protection devices against asynchronous beam dumps are shown in Figure 14-6: the TCDS is a fixed absorber that protects the downstream extraction septum MSD, and the TCDQ is a movable absorber that, together with the secondary collimator TCS, protects the superconducting quadrupole Q4 and further downstream elements, including the arc. The increased beam intensity and brightness for the HL-LHC requires a redesign of the TCDS and TCDQ absorbers.

An upgraded TCDQ will already be installed before LHC Run 2. The new design, which is described in detail in Ref. [10], includes an extension of the absorber length from 6 m to 9 m, and the replacement of the higher density graphite absorber material by different grades (1.4 g/cm$^3$ and 1.8 g/cm$^3$) of carbon fibre composites (CfC). The energy deposition and induced thermal stresses then remain acceptable during an asynchronous abort and the protection of Q4 and downstream elements remain sufficient, with a maximum energy density in the magnet coils of around 20 J/cm$^3$. The TCS collimators will be upgraded with integrated button BPMs in the jaws, which should allow faster and more accurate setup.

A similar level of redesign will be needed for the TCDS absorber. There is one TCDS system, currently consisting of two units per beam. The robustness of the present TCDS and the protection of the MSD magnets in the case of an asynchronous beam dump with full intensity HL-LHC beams need to be verified, and the absorber material and/or length need to be adapted if necessary. Any additional length will slightly reduce the aperture for the circulating or extracted beams by a small fraction of a sigma, which should be acceptable.

A further upgrade of the TCDQ is not part of the HL-LHC baseline; an upgrade of the TCDS is.

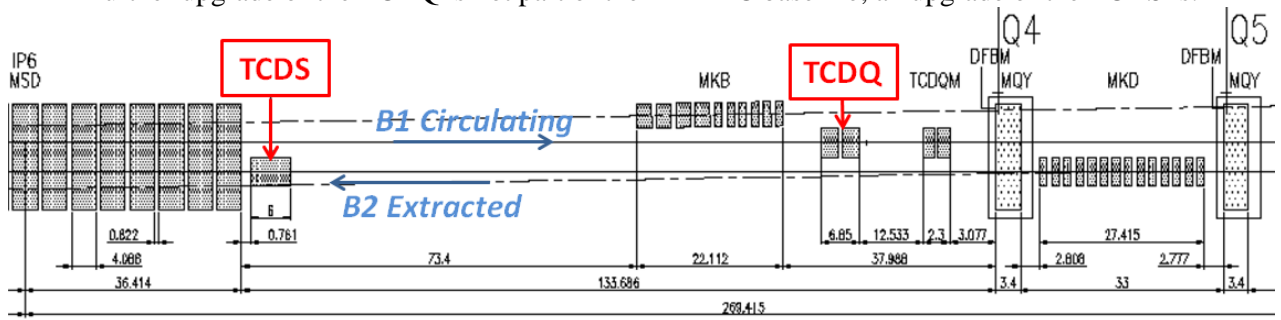

Figure 14-6: Schematic layout of the beam dump area right of P6 as used during LHC Run 1, showing the extraction absorber element TCDS on Beam 2 and the TCDQ on Beam 1.